\documentclass[twocolumn,showpacs,preprintnumbers,amsmath,amssymb,aps,prl]{revtex4}

\usepackage{amssymb,amsfonts,amsmath}
\usepackage{euscript}
\usepackage{graphics}
\usepackage{setspace}
\usepackage{graphicx}
\usepackage{setspace}
\usepackage{dcolumn}
\usepackage{bm}

\begin{document}

\preprint{}

\title{Collective Chemotactic Dynamics in the Presence of Self-Generated Fluid Flows}

\author{}
 \affiliation{}
\author{Enkeleida Lushi$^{1,2}$, Raymond E. Goldstein$^{3}$, Michael J. Shelley$^1$}
\affiliation{$^1$Courant Institute of Mathematical Sciences, New York University, NY, USA\\
$^2$Department of Mathematics, Imperial College London, London, UK\\
$^3$Department of Applied Mathematics and Theoretical Physics, University of Cambridge, Cambridge, UK}


\begin{abstract}
In micro-swimmer suspensions 
locomotion necessarily generates fluid motion, and it is known that such flows can
lead to collective behavior from unbiased swimming. We examine
the complementary problem of how chemotaxis is affected by self-generated flows.
A kinetic theory coupling run-and-tumble chemotaxis to the  
flows of collective swimming shows separate branches of chemotactic and
hydrodynamic instabilities for isotropic suspensions, the first driving
aggregation, the second producing increased orientational order in
suspensions of ``pushers'' and maximal disorder in suspensions of ``pullers''.  
Nonlinear simulations show that hydrodynamic interactions can limit and modify chemotactically-driven 
aggregation dynamics. In puller suspensions the dynamics form aggregates that are 
mutually-repelling due to the non-trivial flows. In pusher suspensions 
chemotactic aggregation can lead to destabilizing 
 flows that fragment the regions of aggregation.
\end{abstract}


\maketitle

A growing body of experimental work has established that suspensions
of motile microorganisms can develop complex large-scale patterns of collective swimming 
at sufficiently high concentration \cite{DombrowskiEtAl04,Ramaswamy}.  
This behavior generally occurs in the {\it absence} of directional cues
for swimming, purely as a consequence of steric and hydrodynamic interactions
between the cells.  Yet, there are many circumstances in which
cells exhibit chemotaxis, directed motion in response to chemical gradients, and this process
by itself can lead to complex spatio-temporal pattern formation \cite{BudreneBerg}.  
As swimmer-generated flows
may also advect any chemoattractant field, it is natural then to ask
how self-generated fluid flows in suspensions of microorganisms affect modes of 
communication \cite{Bassler} and aggregation.
Here we present an analysis of this issue and suggest potential realizations of this
pattern-forming system.

Chemotactic focusing of cell concentration has been
studied using the classical Keller-Segel (KS)
model \cite{KellerSegel71a} and in theories incorporating the 
run-and-tumble (RT) \cite{Berg83} dynamics of bacterial
motion \cite{BearonPedley00,ChenEtAl03}.
We extend a well-known kinetic model for modulated RT
dynamics to include flows produced by the active stresses
due to swimming. A simpler version of our model is considered in
\cite{BearonPedley00} to study swimmer transport and rotation in 
a given background shear flow. Without RT dynamics, our model
reduces to one for active suspensions \cite{SaintShelley08} which captures 
the large-scale flows seen in experiments 
\cite{DombrowskiEtAl04,CKGG11} and illuminates
the effect of propulsion mechanism ({\it pusher} vs. {\it puller}) on
large-scale dynamics and stability.  When swimmers produce a
chemoattractant leading to aggregation, the self-generated flows can
have a large effect; pushers create complex flows that can bound
growth in organism density, while pullers show limited pattern
coarsening and isolated aggreggates repelling due to non-trivial
flows.  Merging the RT and active suspension models is seamless as
both are kinetic theories with conformation variables the particle
position and orientation \cite{KelaThesis}.

Consider a suspension of swimmers at local concentration
$\Phi(\mathbf{x},t)$, each of which moves at a constant speed $U$ in a
run-and-tumble dynamics.  They move in a fluid of local velocity ${\bf
u}({\bf x},t)$ and produce a chemoattractant of concentration
$C(\mathbf{x},t)$ and molecular diffusivity $D_C$.  We choose
rescalings based on the swimmer contribution to the fluid stress
tensor (below), with a characteristic length $\ell=l/\phi$, where $l$
is the swimmer size and $\phi\equiv l^3\langle \Phi \rangle$ is the
effective mean volume fraction in suspension.  Scaling time by
$\ell/U$, $C$ evolves as
\begin{eqnarray}\label{chemo}
 C_t + \mathbf{u} \cdot {\bm\nabla} C = Pe^{-1} \nabla^2 C -\beta_1 C +\beta_2 \Phi~,
\end{eqnarray}
where $\beta_1$ and $\beta_2$ are rate constants for chemoattractant 
self-degradation and production, and the P{\'e}clet number
$Pe=U\ell/D_C$ measures the strength of diffusion to
advection on the intrinsic scale $\ell$.  Collective swimming may generate coherence on 
larger scales with higher speeds, increasing the importance of advection.  
Without advection this is the KS model \cite{KellerSegel71a}.
In the case of {\it E. coli} \cite{Saragosti} gives $U\sim 25$ $\mu$m/s,
$l\sim 5$ $\mu$m, $\phi\sim 0.1$ at a cell concentration of $10^9$ cm$^{-3}$, so 
$\ell\sim 50$ $\mu$m.  With  $D_C\sim 5\times 10^{-6}$ cm$^{2}$/s we obtain 
$Pe\simeq 2.5$, $\beta_1 \simeq 
0.008$ and $\beta_2 \simeq 0.004$.  For faster-swimming organisms, such as
marine bacteria \cite{Stocker}, this intrinsic P{\'e}clet number can reach ${\cal O}(10-20)$.

\begin{figure*}
\centering
\includegraphics[width=1.9\columnwidth]{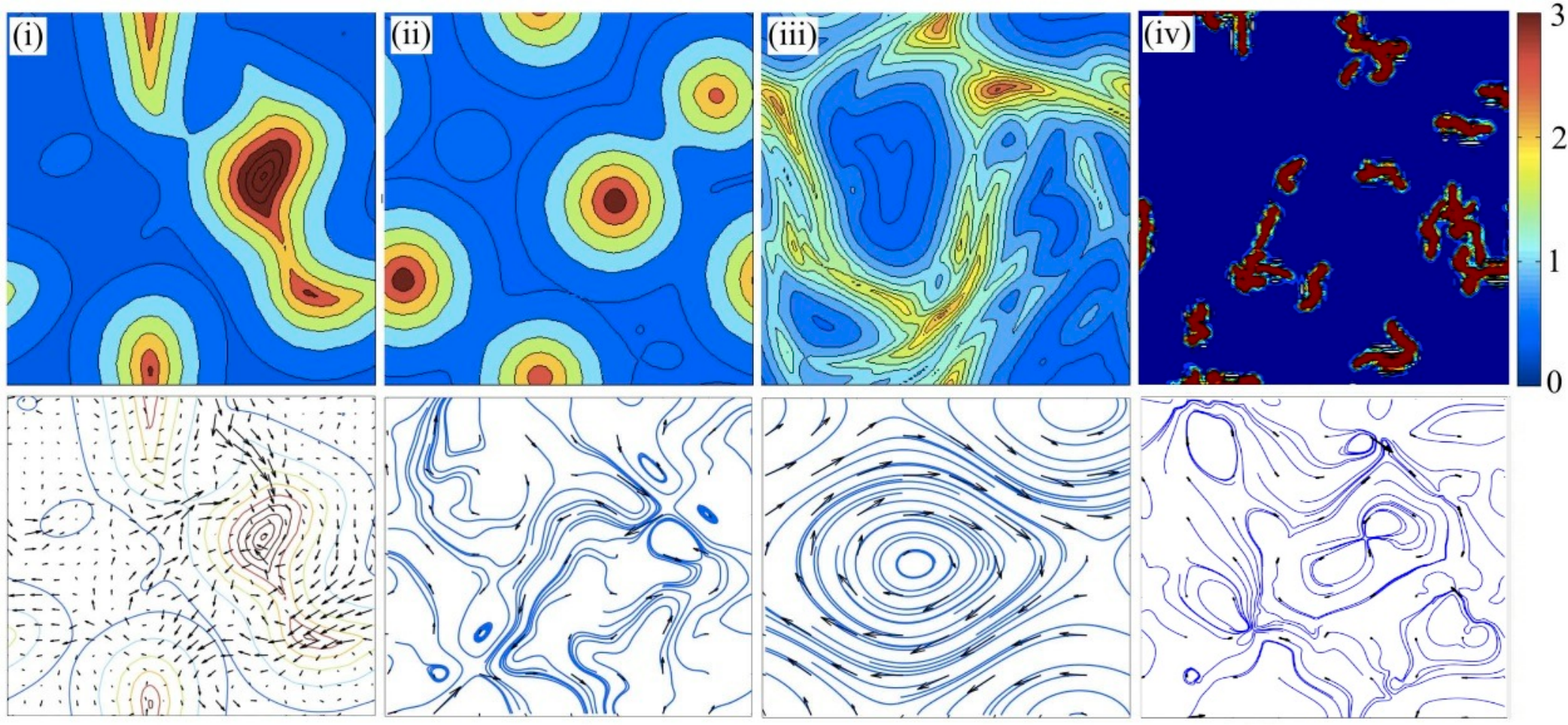}
 \caption{(color online).  Chemotactic suspensions at large times, $t=3000$.  Swimmer concentration
$\Phi$ (top) and mean direction ${\bf n}$ (bottom) for (i) chemotactic neutral
swimmers ($\mathbf{u}=0$), and (ii) $\Phi$ (top), fluid streamlines and velocity field ${\bf u}$ (bottom) for chemotactic pullers. 
(iii,iv) $\Phi$ (top), streamlines and ${\bf u}$ (bottom) for chemotactic
pushers. Parameters $\beta_1=\beta_2=0.25$, $Pe=20$, $\gamma=1$, $D_T=D_R=0.025$. 
Parameters $\lambda_0$ and $\chi$ are indicated in Figs. 2c,d for cases (i-iii) and are $\lambda_0=5, \chi=0.6$ for case (iv). 
See \cite{SupplMovies} for movies of swimmer concentration dynamics.}
\vspace{-0.15in}
 \label{fig1}
\end{figure*}

The configuration of swimmers is given by a distribution function
$\Psi(\mathbf{x},\mathbf{p},t)$ of the center of mass position
$\mathbf{x}$ and orientation $\mathbf{p}$ satisfying the
Fokker-Planck equation
\begin{align}\label{runandtumble3D}
 \Psi_t=&- \left[ \lambda(\mathcal{D}_t C) \Psi - \frac{1}{4 \pi}
\int\!d \mathbf{p'} \lambda(\mathcal{D}_t C) \Psi(\mathbf{x},\mathbf{p'},t)   \right] \nonumber \\
 & - {\bm\nabla}_{\bf x} \cdot ( \Psi \dot{\mathbf{x}} ) - {\bm\nabla}_{\bf p} \cdot ( \Psi \dot{ \mathbf{p}})~,
\end{align}
where the local swimmer concentration is
$\Phi(\mathbf{x},t)=\int\!d\mathbf{p}\Psi(\mathbf{x},\mathbf{p},t)$.
The bracketed term in (\ref{runandtumble3D}) describes the
effect of RT chemotaxis based on a swimming dynamics of straight runs
and modulated reorientations (tumbles) where $\lambda(\mathcal{D}_t
C)$ is a tumbling frequency, the probability of a bacterial
tumble event as a function of the chemoattractant temporal gradient
 $\mathcal{D}_t C = C_t+(\mathbf{p}+ \mathbf{u})
\cdot {\bm\nabla} C$ along a swimmer's path.  Experiments
\cite{MacnabKoshland72} show that when $\mathcal{D}_t C>0$ the
tumbling frequency is reduced, and is otherwise constant, as captured by the
biphasic form
$\lambda(\mathcal{D}_t C)=\lambda_0 \max(\min(1-\chi \mathcal{D}_t
C,1),0)$, a linearized version of an earlier model \cite{ChenEtAl03, Saragosti}.  
The fluxes in (\ref{runandtumble3D}) are
\begin{eqnarray}
\dot{\mathbf{x}} =  \mathbf{p} + \mathbf{u}, \hspace{0.2in} 
\dot{ \mathbf{p}} = (\mathbf{I}- \mathbf{p}\mathbf{p}) 
\left(\gamma{\bf E}+{\bf W}\right)\mathbf{p}. \label{pdot}
\end{eqnarray}
The particle velocity $\dot{\mathbf{x}}$ includes swimming at
constant speed (non-dimensionalized to unity) in the axis direction
$\mathbf{p}$ ($|\mathbf{p}|=1$), translation by the fluid
velocity $\mathbf{u}$. (In the KS model \cite{KellerSegel71a},
swimmer speed is linear in the chemical gradient.)  The angular
velocity $\dot{ \mathbf{p}}$ follows Jeffrey's equation \cite{Jeffrey22} 
where ${\bf E}=({\bm\nabla}_{\bf x}{\bf u}+{\bm\nabla}_{\bf x}{\bf u}^T)/2$
is the rate-of-strain tensor, ${\bf W}=({\bm\nabla}_{\bf x}{\bf
u}-{\bm\nabla}_{\bf x}{\bf u}^T)/2$ is the vorticity tensor. 
For rod-like swimmers, the shape factor is $\gamma\sim 1$. 

The fluid velocity $\mathbf{u}(\mathbf{x},t)$ produced by the suspension
satisfies the Stokes equations driven by an ``active" stress
$\mathbf{\Sigma}^a$ arising from particle locomotion:
\begin{eqnarray}
-&\nabla_{\bf x}^2 \mathbf{u} +{\bm\nabla}_{\bf x} q = {\bm\nabla}_{\bf x} \cdot \mathbf{\Sigma}^a,
 \hspace{0.2in} 
{\bm\nabla}_{\bf x} \cdot \mathbf{u} = 0 \label{Stokes}\\
 &\mathbf{\Sigma}^a (\mathbf{x},t) = 
\alpha \int\!d\mathbf{p} \Psi (\mathbf{x},\mathbf{p},t) 
\mathbf{pp}. 
\label{stress_nondim}
\end{eqnarray}
The active stress is an orientational average of the force dipoles 
$\alpha\mathbf{pp}$ the cells exert on the fluid \cite{SaintShelley08}, 
where $\alpha$ is an $\mathcal{O}(1)$ constant by our rescaling. A
cell that self-propels by front-actuation (a puller) has stresslet strength 
$\alpha>0$, and a rear-actuated cell (pusher) has $\alpha<0$. The case of 
``neutral" cells ($\alpha={\bf u}=0$) is the closest this model approaches the
KS \cite{KellerSegel71a} and RT models \cite{ChenEtAl03, Saragosti}.

\begin{figure*}[t]
\includegraphics[width=1.9\columnwidth]{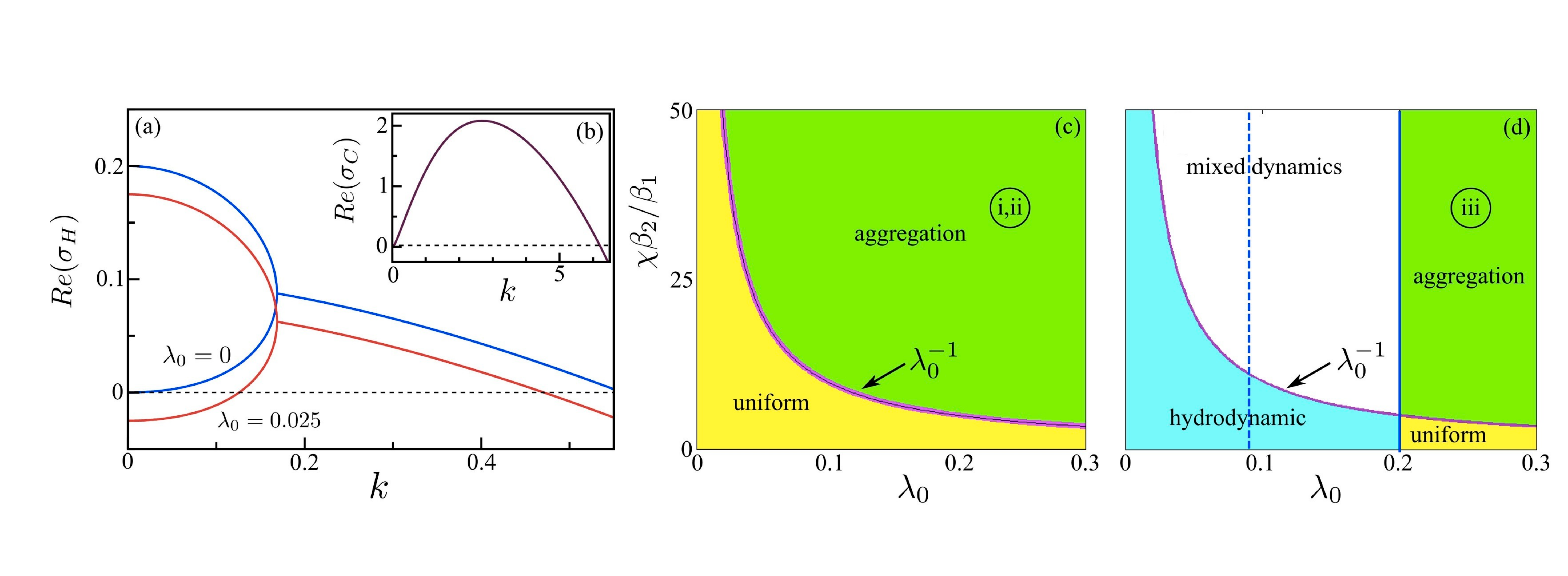}
\label{stab}
\vskip -1.0cm
\caption{(color online).  Linear stability analysis. (a) Branches of the 
hydrodynamic instability, with and without tumbling, for pushers.  
(b) Chemotactic branch for $\chi=35$, $\lambda_0=0.25$, 
$\beta_1=\beta_2=1/4$ and $Pe=20$. Regimes diagram for (c) neutral 
swimmers and pullers, and (d) pushers. Solid curves give linear stability 
boundaries for long waves. Dashed lines show shifted boundaries in nonlinear 
simulations at finite box size. Encircled labels (i-iii) denote parameters 
used in simulations (Fig. 1a-c).}
\vspace{-0.15in}
\end{figure*}

We first illustrate the effect hydrodynamics has on aggregation. 
From nonlinear simulations of Eqs.~(1-6), Fig.~1 shows the swimmer 
concentration $\Phi(\mathbf{x},t)$ and mean orientation 
${\bf n}=\int d {\bf p}\Psi/\Phi$ at late times, having started near uniform 
isotropy, for neutral,  puller, and pusher suspensions. All share a dominant
self-aggregation instability, but differing (or no) hydrodynamic
interactions. Neutral swimmers show aggregation and pattern
coarsening. Pullers show limited aggregation into circular spots 
 kept apart by non-trivial fluid flows. Pushers create
complex fluid flows and fragmented aggregation regions. 

These behaviors can be understood through a stability analysis of 
uniform isotropic suspensions. For simplicity,
consider rod-like ($\gamma=1$) swimmers and a
quasi-static chemoattractant field 
$Pe^{-1} \nabla^2 C-\beta_1 C =-\beta_2 \Phi$,
which slaves $C$ to $\Phi$.  The tumbling frequency is simplified to
$\lambda(\mathbf{p}) = \lambda_0\left(1- \chi \mathbf{p} \cdot \nabla
C \right)$.  A steady state is $\Psi_0=1/4\pi$ ($\Phi=1$),
${\bf u}={\bf 0}$, and $C_0=\beta_1/\beta_2$.  
Perturbations of the form
$\epsilon(\tilde{\Psi}(\mathbf{p},\mathbf{k}),\tilde{C}(\mathbf{k}))
\exp (i \mathbf{k} \cdot \mathbf{x} + \sigma t)$,
yield 
\begin{align}\label{PsiLinDecomp}
(\sigma + \lambda_0 + i \mathbf{k} \cdot \mathbf{p} ) 
\tilde{\Psi}&= \frac{\lambda_0}{4 \pi} 
\left( \frac{ik\chi \beta_2 (\hat{\mathbf{k}} \cdot \mathbf{p})}
{\beta_1 + k^2 Pe^{-1}} + 1  \right) \tilde{\Phi} 
\nonumber \\
&-\frac{3 \alpha}{4 \pi}(\hat{\mathbf{k}} \cdot \mathbf{p}) 
\mathbf{p} \cdot (\mathbf{I}-\hat{\mathbf{k}}\hat{\mathbf{k}}) 
\tilde{{\bf\Sigma}}^p \mathbf{k}~,
\end{align} 
where $\tilde{\Phi}=\int\!d\mathbf{p'}\tilde{\Psi}'$  and 
$\mathbf{k}=k\hat{\mathbf{k}}$. Since $\mathbf{\tilde{\Sigma}}^p=\int\!d\mathbf{p'}
\tilde{\Psi}'\mathbf{p'p'}$, this
is a linear eigenvalue problem for $\tilde{\Psi}$ and
$\sigma$. The first term on the RHS is chemotactic (C) and has
unstable dynamics restricted to the zeroeth azimuthal mode on
$|\mathbf{p}|=1$. The second is hydrodynamic (H), with
unstable dynamics restricted to the first azimuthal mode. This yields \textit{uncoupled}
relations for growth rates $\sigma_{C,H}$,
\begin{eqnarray}
\frac{2}{\lambda_0} &=&R\left[ 2+a_C\log \left(  \frac{a_C-1}{a_C+1}\right) \right] 
-\frac{1}{ik}\log \left(  \frac{a_C-1}{a_C+1}\right)\nonumber \\
\frac{4 k}{3 i \alpha} &=& 2a_H^3 -\frac{4}{3}a_H + (a_H^4-a_H^2)
\log \left(\frac{a_H-1}{a_H+1}\right)~,\label{chem-inst}
\end{eqnarray}
where 
$a_{C,H}=(\sigma_{C,H} + \lambda_0)/ik$ and $R=\chi
\beta_2/(\beta_1+k^2/Pe)$.
We refer to these as the {\it
chemotactic} and {\it hydrodynamic} relations, respectively. The first
induces growth in concentration fluctuations, while the second
increases orientational order. The two are coupled only through the
basal tumbling rate $\lambda_0$ which in the hydrodynamic relation
only shifts the growth rate. Further, the chemotactic instability gives
rise to normal stresses of the form $\mathbf{\tilde{\Sigma}}^p=
\hat{\mathbf{k}}\hat{\mathbf{k}} -
\hat{\mathbf{k}}_\perp \hat{\mathbf{k}}_\perp$, while the hydrodynamic
instability gives shear stresses of the form
$\mathbf{\tilde{\Sigma}}^p= \hat{\mathbf{k}}\hat{\mathbf{k}}_\perp +
\hat{\mathbf{k}}_\perp \hat{\mathbf{k}}$.

For pushers ($\alpha<0$) the hydrodynamic instability has a
finite bandwidth (Fig.~2a), though with maximal
growth rates at $k=0$. Tumbling shifts down the $Re(\sigma_H(k))$
branch by $\lambda_0$ for all $k$, further stabilizing the system.
Long-wave asymptotics of (\ref{chem-inst}) give two solution
branches: $\sigma_{H1}\simeq -\alpha/5-\lambda_0 +15/7\alpha k^2$
and $\sigma_{H2}\simeq -\lambda_0+O(-\alpha k^2)$. There is no
hydrodynamic instability for pullers \cite{SaintShelley08}.  
Fig.~2b shows the chemotactic growth rate.
Small $k$ asymptotics yields
$\sigma_C \approx k^2/(3\lambda_0) [(\chi \beta_2/\beta_1) \lambda_0
-1]$: for $(\chi \beta_2/\beta_1) >1/\lambda_0 $ there
are wavenumbers with $Re(\sigma_C(k))>0$, shown in one case as a
finite band of unstable modes whose width is controlled by
chemo-attractant diffusion.

From Fig.~2a, we can obtain a range for $\lambda_0$ for which there is
a hydrodynamic instability in pusher suspensions.  Heuristically, $\lambda_0$ 
sets an effective rotational diffusivity, and $\lambda_0 \geq 0.2$
turns off the hydrodynamic instability for any system size. For 
$L=50$ and the diffusion constants used in simulations, $\lambda_0
\geq 0.09$ suffices. This information is assembled in Fig. 2~c,d as phase
diagrams that relate the parameters to various dynamical regimes.

Numerical studies of the full nonlinear dynamics (\ref{chemo}--\ref{stress_nondim}) 
were done in 2D, with a box size $L=50$  large enough to include several 
unstable linear modes. Swimmer translational and rotational diffusions are 
added in the model to control the growth of steep gradients over long-times. An
initial random perturbation of the uniform isotropic state is used:
$\Psi(\mathbf{x},\mathbf{p},0) = 1/2\pi +\Sigma_i \epsilon_i \cos(\mathbf{k}_i \cdot
\mathbf{x}+\xi_i)Q_i(\mathbf{p}_i) $ with random coefficients
$|\epsilon_i |<0.01$, $\xi_i$ an arbitrary phase and $Q_i$ a low-order
polynomial. The initial chemo-attractant concentration is uniform with
$C(\mathbf{x},0)=\beta_1/\beta_2=1$. Figure 1 shows long-time swimmer
concentration $\Phi$ for four illustrative cases. In each case, 
concentration $C$ closely tracks $\Phi$. Cases (i-iii) share
the same chemotactic instability, but differ in swimming actuation:
$\alpha=0,1,-1$.
 
The expected regimes of these three cases are shown in Figs.~2c,d.
For neutral swimmers, aggregation dominates and the dynamics is
typified by the formation of a few regions of steadily increasing
concentration that slowly coarsen (Fig. 1a).  The maximum swimmer
concentration (Fig. 3) shows little sign of the rapid self-focussing
associated with finite-time chemotactic collapse
\cite{ChildressPercus81} of the KS model, which here may be due to the
fixed swimming speed \cite{SchnitBerg90}.  While the initial
aggregation for pullers (Fig.~1b) is similar to that for neutral
swimmers (Fig. 3) its long-time behavior is very
different. Concentration growth and coarsening cease as the dynamics
enters a near steady-state with circular regions of high
concentration. Active-stress driven flows suppress further coarsening
by pushing nearby peaks apart and apparently maintain the few
remaining high concentration regions.

\begin{figure}[t]
\includegraphics[width=0.7\columnwidth]{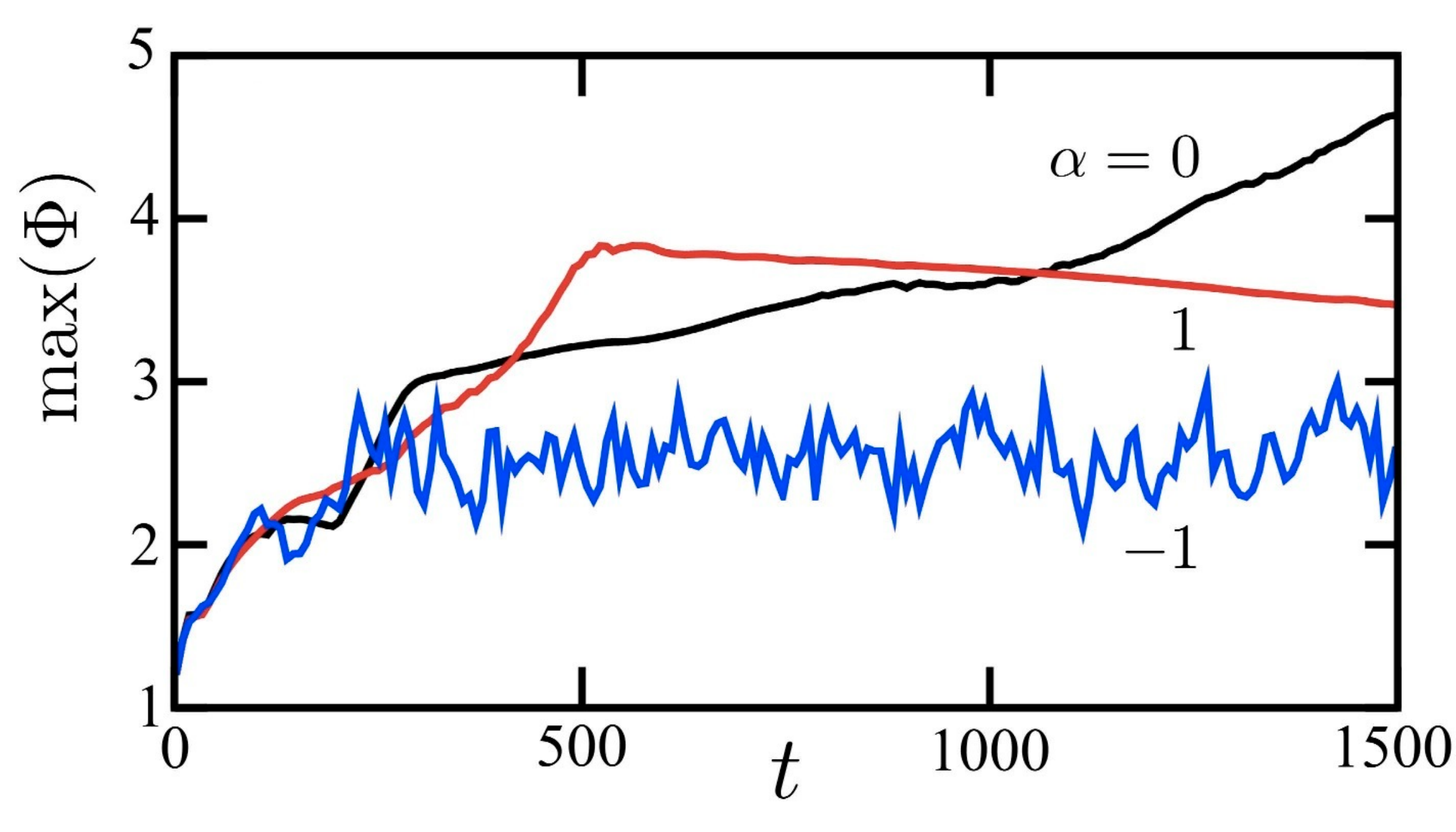}
  \label{figure3}
  \vspace{-0.1in}
     \caption{(color online) Measures of growth: maximum swimmer concentration 
     for cases (i-iii) in Fig. 1. }
\vspace{-0.2in}
\end{figure}

For pushers (Fig.~1c), linear theory gives only a chemotactic
instability, and the dynamics is indeed initially dominated by
aggregation as is evidenced by the early rapid growth of normal
stresses relative to shear stresses (not shown). However, aggregation
into a regions with high swimmer concentration creates a destabilizing
active stress, giving rise to unsteady fluid flows. These flows
fragment the peaks while pushing them around the domain. The dynamics
is one of constant aggregation and flow instability, which apparently
suppresses further growth in swimmer concentration (Fig. 3).

Lastly, we examine in Fig.~1(iv) the dynamics that arises with parameters
close to those measured by Saragosti {\it et al} \cite{Saragosti}
(before our rescaling) in their experiments of {\it E. coli}
chemotaxis.  These parameters lie far to the right of the aggregation
regime of Fig. 2d as $\lambda_0$ is $20$ times higher than at the
predicted threshold for suppressing hydrodynamic instabilities. Not
surprisingly, the simulations show chemotactic aggregation into very
high peaks. Once the swimmer concentration in those peaks is large
enough, the active stresses give rise to small-scale and localized
fluid flows (cf. Fig.~1(iii)). These local flows do appear to be
implicated in the slow ``wriggling'' we observe of the saturated
aggregates (see Supplementary Material \cite{SupplMovies}).  
The experiments of Saragosti {\it et al} \cite{Saragosti}, which
are performed in confined micro-channels and capillaries, show instead the
development of traveling concentration waves of chemotactic bacteria. These
traveling waves were initiated in the experiments through an initial
concentration by centrifugation of the swimmer population to one end of the
channel. We do not observe the spontaneous formation of such traveling waves
here though ours is an open system (though confined geometrically by the
assumed periodicity length) and the initial swimmer state is un-oriented and
nearly homogeneous. The combined effects of a confining geometry and the
initial concentration of swimmers has yet to be examined in our theoretical
system.

We have shown that the intrinsically generated fluid flows arising from
collective swimming of microorganisms can modify patterns of chemotactic
aggregation and, most importantly, can limit aggregate concentration. This is
unlike chemotactic models that predict concentration blow-up or include
artificial terms to cap growth. While we have emphasized hydrodynamic effects
in attractive chemotactic dynamics, it is important to remember that ours is
a dilute to semi-dilute theory that does not capture near-interactions
between swimmers, hydrodynamic or otherwise. In denser suspensions swimmer
size limits local swimmer density through steric interactions though as yet
well-founded models that combine these with hydrodynamic interactions do not
exist. Nonetheless, we expect similar results when large-scale coherence is
driven by steric effects \cite{CKGG11, DrescherEtAl11}. On that note,  
steric effects with no hydrodynamics may also limit aggregation of
chemotactic random walkers \cite{Taktikos12}.

Finally, these auto-chemotactic effects can be seen as complementary to the enhanced
mixing by swimmers \cite{KelaThesis} that has also been explored for
microfluidic applications \cite{KimBreuer07}.  Systematic studies of
the interplay between chemotaxis and locomotion-generated fluid flow
should be possible through controlled introduction of exogeneous
chemoattractants to trigger aggregation, through the interplay of
quorum sensing and chemotaxis \cite{Park}, and perhaps by specific
genetic engineering of the dynamics of locomotion and chemosensing
\cite{Liu}.

This work was supported in part by NSF grants DMS-0652775 and
DMS-0652795, and DOE grant DEFG02-00ER25053 (E.L., M.J.S.) and an ERC
Advanced Investigator Grant 247333 (R.E.G.).


\begin{thebibliography}{26}

\bibitem{DombrowskiEtAl04}
C.~Dombrowski, {\it et al.}, \newblock{Phys. Rev. Lett.} {\bf 93}, 098103 (2004); 
L.H. Cisneros, {\it et al.}, Exp. Fluids {\bf 43}, 737 (2007).

\bibitem{Ramaswamy} S. Ramaswamy, Ann. Rev. Condens. Matter Phys. {\bf 1}, 323 (2010).

\bibitem{BudreneBerg} E.O. Budrene and H.C. Berg, Nature {\bf 349}, 630 (1991).

\bibitem{Bassler} B.L. Bassler, Cell {\bf 109}, 421 (2002).

\bibitem{KellerSegel71a}
E.F.~Keller, L.A.~Segel, J. Theor. Biol. {\bf 30}, 225 (1971).

\bibitem{Berg83}
H.C.~Berg, {\em Random Walks in Biology} Princeton University Press, 1993

\bibitem{BearonPedley00}
R.N.~Bearon, T.J.~Pedley, Bull. Math. Bio. {\bf 62}, 775 (2000).

\bibitem{ChenEtAl03}
K.C.~Chen, {\it et al.}, J. Math. Bio. {\bf 47}, 518 (2003).

\bibitem{SaintShelley08}
D.~Saintillan and M.J.~Shelley,
Phys. Rev. Lett. {\bf 100}, 178103 (2008); also Phys. Fluids {\bf 20}, 123304 (2008).

\bibitem{CKGG11} L.H. Cisneros, {\it et al.}, Phys. Rev. E
{\bf 83}, 061907 (2011).

\bibitem{KelaThesis} 
E.~Lushi and M.J.~Shelley, preprint (2012).

\bibitem{Saragosti} J. Saragosti, {\it et al.}, Proc. Natl. Acad. Sci. USA {\bf 108}, 16235 (2011).

\bibitem{Stocker} R. Stocker, {\it et al.}, Proc. Natl. Acad. Sci. USA {\bf  105}, 4209 (2008).

\bibitem{MacnabKoshland72}
R.M.~Macnab and D.E.~Koshland, Proc. Natl. Acad. Sci. USA {\bf 69}, 2509 (1972);
N.~Mittal, {\it et al.},
Proc. Natl. Acad. Sci. USA {\bf 100}, 13259 (2003).

\bibitem{Jeffrey22}
G.B.~Jeffery, Proc. R. Soc. London, Ser. A {\bf 102}, 161 (1922).

\bibitem{SupplMovies}
See Supplementary Material at [URL to be inserted by publisher] for movies of the swimmer concentration.

\bibitem{ChildressPercus81}
S.~Childress and J.K.~Percus, Math. BioSci. {\bf 56}, 217 (1981).

\bibitem{SchnitBerg90}
M.J.~Schnitzer, {\it et al.},
Symp. Soc. Gen. Microbiol {\bf 46}, 15 (1990).


\bibitem{DrescherEtAl11}
K. Drescher, {\it et al.},
Proc. Natl. Acad. Sci. USA {\bf 108}, 10940 (2011).

\bibitem{Taktikos12}
J.~Taktikos, {\it et al.},
Phys. Rev. E {\bf 85}, 051901 (2012).

\bibitem{KimBreuer07}
M.J.~Kim and K.S.~Breuer,
Anal. Chem. {\bf 79}, 955 (2007).

\bibitem{Park} S. Park, {\it et al.}. Science {\bf 301}, 188 (2003).

\bibitem{Liu} C. Liu, {\it et al.}. Science {\bf  334}, 238  (2011).

\end{thebibliography}
\end{document}